# MICROMACHINED POLYCRYSTALLINE SiGe-BASED THERMOPILES FOR MICROPOWER GENERATION ON HUMAN BODY

Z. Wang[1,2], V. Leonov[1], P. Fiorini[1], and C. Van Hoof[1]

[1]IMEC vzw, Kapeldreef 75, B-3001, Leuven, Belgium

[2]Catholic University Leuven, B-3000, Leuven, Belgium

## ABSTRACT

This paper presents a polycrystalline silicon germanium (poly-SiGe) thermopile specially designed for thermoelectric generators used on human body. Both the design of the single thermocouple and the arrangement of the thermocouple array have been described. A rim structure has been introduced in order to increase the temperature difference across the thermocouple junctions. The modeling of the thermocouple and the thermopile has been performed analytically and numerically. An output power of about 1 µW at an output voltage of more than 1 V is expected from the current design of thermopiles in a watch-size generator. The key material properties of the poly-SiGe have been measured. The thermopile has been fabricated and tested. Experimental results clearly demonstrate the advantage of the rim structure in increasing output voltage. In presence of forced convection, the output voltage of a non-released thermopile can increase from about 53 mV/K/cm$^2$ to about 130 mV/K/cm$^2$ after the rim structure is formed. A larger output voltage from the thermopile is expected upon process completion.

## 1. INTRODUCTION

Body area networks, consisting of multiple sensors, transducers and transceivers deployed on human body for medical use, call for long-lasting, if possible everlasting, power supplies. Thermoelectric generators (TEGs) transforming the heat of the human body into electrical energy, could provide power autonomy to the nodes of the network and replace traditional batteries. An autonomous body area network powered by TEGs has already been reported [1]. In that application the TEG is based on commercial bismuth telluride (BiTe) thermopiles, which are fabricated by standard machining and serial production techniques. Therefore, their cost is prohibitive and their size is relatively large.

Micromachining technology allows the miniaturization of the traditional thermopiles. Till now, several miniaturized thermopiles have been reported, either based on BiTe or on poly-SiGe [2-6]. Besides the merit of providing power autonomy, miniaturized thermopiles are cheaper, smaller and can be monolithically integrated with other components, such as power management circuits.

However, till now, none of the current miniaturized thermopiles can provide a usable power output when applied on human body due to its adverse thermal conditions，namely the low temperature differences from the ambient air and the large thermal resistance of the body. This paper presents a type of poly-SiGe based micromachined thermopile specially designed for human body applications.

## 2. DEVICE PRINCIPLE

Fig. 1(a) shows the scheme of a typical TEG, which consists of a number of thermocouples connected thermally in parallel and electrically in series, sandwiched between a hot plate and a cold plate. Fig. 1(b) shows the equivalent thermal circuit describing a TEG placed on human body.

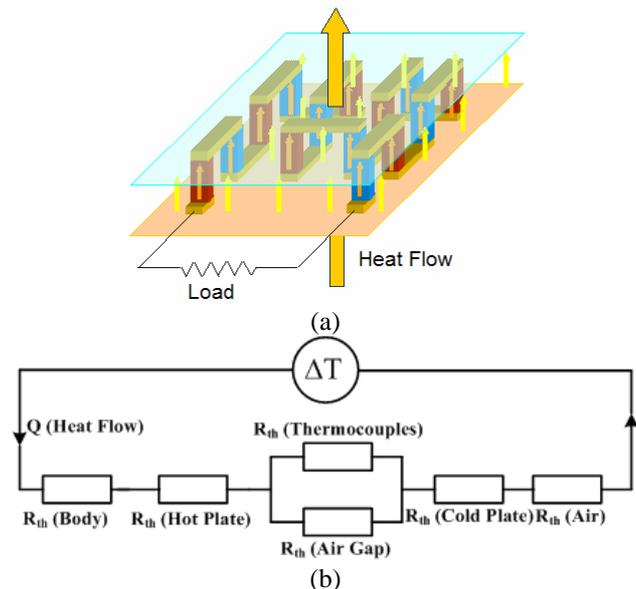

Fig.1. (a). Scheme of a typical TEG; (b). Equivalent thermal circuit of a TEG placed on human body.

Normally the thermal resistances of human body and ambient air are much larger than that of the thermopile so that the heat flow can be considered constant and approximately independent of the specific design of the thermopile. Under this hypothesis it can be easily shown that the maximum power output is obtained when the heat flow through the thermocouples is equal to that through the air gap, namely when the thermal resistance of the thermocouples is equal to that of the air gap.

If the thermopile of Fig. 1 is fabricated by micromachining techniques, both the lateral size and the height of the single thermocouple will be of the order of a few micrometers. Thermocouples can be fabricated on a silicon substrate and capped by a silicon top plate forming the structure shown in Fig. 2a. As the thermoelectric materials have a thermal conductivity about 100 times larger than the one of air, the condition of maximum power (matched thermal resistance of air gap and thermoelectric material) corresponds to the situation where the thermocouples occupy only about 1/100 of the hot plate area. (For clarity this feature is not represented correctly in the figure). The total thermal conductance of the device can be roughly evaluated as twice the one of the air between the plates. As they are very close (several micrometers) the thermal conductance is large leading to negligible temperature difference and output power. Furthermore, as the temperature difference is low, a large number of thermocouples connected electrically in series are needed to generate a usable voltage output, such as 1 V.

An alternative and more efficient arrangement is shown in Fig. 2(b). Only the part of the Si wafer containing the thermocouples is used and mounted between metal cold and hot plates. The thermal resistance of the air gap is now hundreds of times larger than the case in Fig. 2(a). Then the temperature drop across the thermocouples and the output voltage are effectively increased. The number of thermocouples necessary to obtain 1 V voltage output also decreases drastically. As the condition of optimum power corresponds to the equality of the thermal resistance of air gap and thermoelectric material, the area occupied by the thermoelectric material further decreases, it can become so small to make the chip handling inconvenient. A solution to this problem consists of using larger chips which are deeply etched where the thermocouples are not placed. Moreover, in order to deal with a more regular chip, the thermopile is arranged in a rim shape, as shown in Fig 2(c). For clarity, the thermocouples arranged in a rim shape placed on the bottom chip are illustrated in a 3D schematic in Fig. 3. In the following we will concentrate on the modeling and fabrication of this structure, which, according to the above discussion, is the one which better combines ease of fabrication and higher performances.

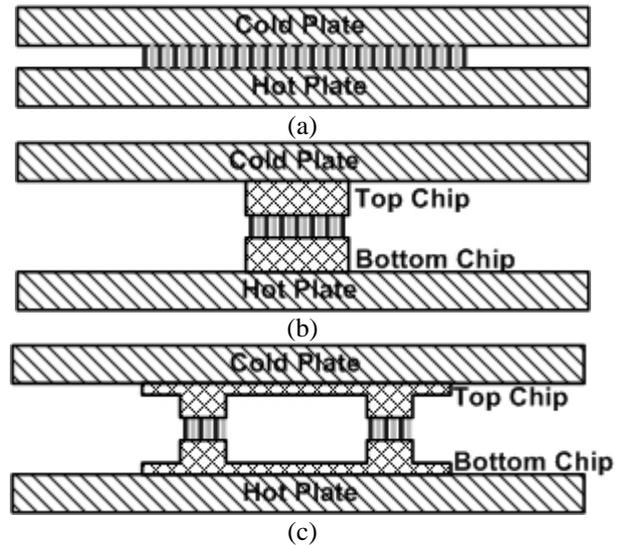

Fig. 2. (a) Cross-sectional view of thermopile sandwiched between a hot plate and a cold; (b) A bottom chip and a top chip are inserted in between the hot plate and the cold plate; (c) The bottom chip and the top chip are enlarged and deeply etched and the thermopile is arranged in a rim shape.

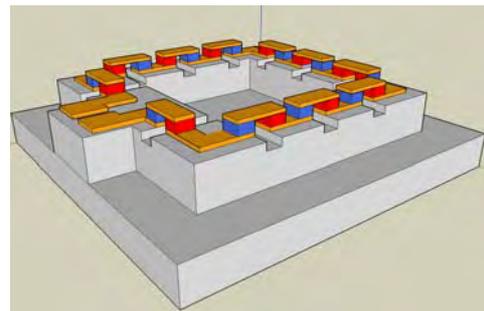

Fig. 3. 3D schematic of the thermocouples in a rim shape.

## 3. MODELING

In this section, the structure and the thermal properties of the single thermocouple are first discussed. Then, based on the result on the single thermocouple, the output performances of the full TEG are computed analytically.

Each single thermocouple is composed of one p- and one n-type poly-SiGe legs, which are interconnected by aluminum pads at the junctions, as shown in Fig 4. If there is a temperature difference between the two junctions, a voltage is built up due to the Seebeck effect. In our design, some measures have been taken to improve the thermal isolation. The thermal isolation of the cold junctions from the hot die surface has been improved by etching a trench into the Si substrate below. A small vertical step, varying from 0.5 to 3 μm in height, has been

Z. Wang[1,2], V. Leonov[1], P. Fiorini[1], and C. Van Hoof[1]

made between the two junctions by means of an additional sacrificial layer. The thermocouple leg has to climb over this vertical step, thus increasing its length and hence its thermal resistance. Moreover, the width of the middle part of the thermocouple legs is reduced to increase the thermal resistance, while the width of the thermocouple legs at junctions remains relatively large to ensure a low contact resistance.

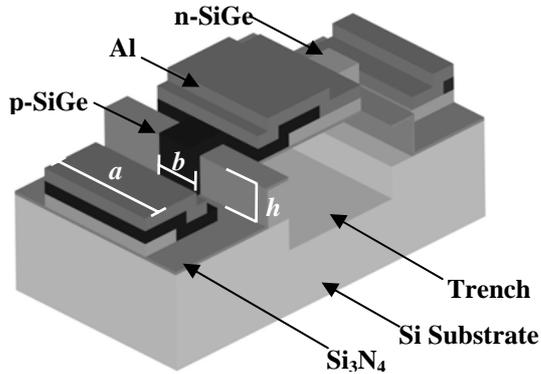

*Fig. 4. 3D schematic drawing of a thermocouple with a 2.5-μm-high step.*

The single thermocouple (Fig. 4) has been modeled numerically. A fully 3D model of a single thermocouple bonded to the top chip (Fig. 10(f)) has been built in FEM software MSC Marc for parametric optimization. The thermal resistance can be determined from FEM simulation. As boundary conditions, a known heat flow is forced into the bottom surface and a fixed temperature is applied to the top surface. As an example of the simulation results, the steady-state temperature distribution is shown in Figure 5. For clarity, only the temperature distribution on the bottom is shown. The thermal resistance of this single thermocouple can be calculated as the ratio between the temperature difference from the bottom surface to the top surface and the heat flow.

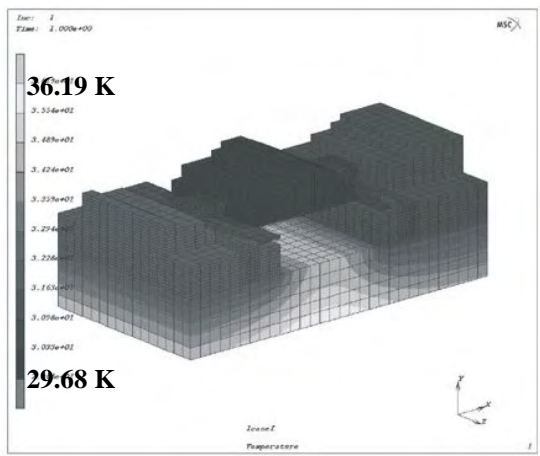

*Fig. 5. FEM simulation of thermocouple in MSC Marc under a fixed heat flux (the top chip is not shown for clarity).*

In the single thermocouple design (Fig. 4), the end width of thermocouple $a$, the middle width of thermocouple $b$ and the thermocouple height $h$ can be separately optimized in order to investigate their influence on the thermal resistance. As shown in Fig. 6, the thermal resistance depends almost linearly on the thermocouple height $h$. The data are obtained with the end width $a$ fixed at 10 μm and the middle width $b$ fixed at 3 μm.

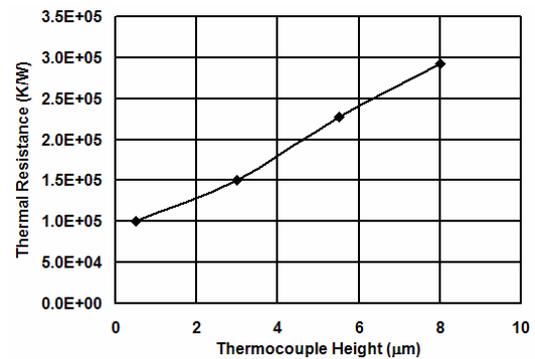

*Fig. 6. Thermal resistance of a single thermocouple depending on the thermocouple height.*

On the other hand, the thermal resistance is inversely proportional to the width when other parameters are fixed. This is shown in Fig. 7, where the thermal resistance decreases from $2.58\times10^5$ K/W to $1.29\times10^5$ K/W when the middle width $b$ increases from 0.5 μm to 4 μm.

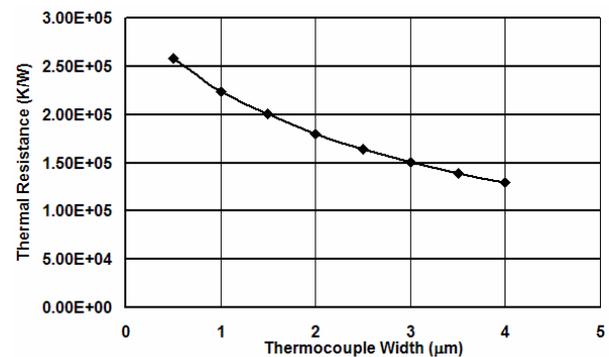

*Fig. 7. Thermal resistance of a single thermocouple depending on the thermocouple width.*

In the final mask layout, various types of thermocouples have been designed by varying the thermocouple widths $a$ and $b$. The end width $a$ varies from 3 μm to 10 μm while the middle width $b$ varies from 1 μm to 3 μm. The thermal resistance of each design type is shown in Fig. 8 for a fixed value of *h* equal to 0.5 μm.

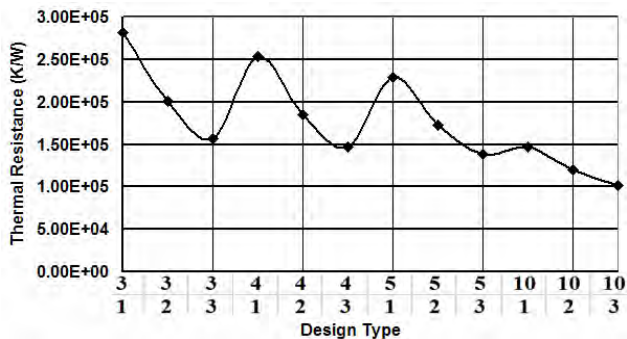

*Fig. 8. Thermal resistance of the thermocouples implemented in the final mask. The two rows of numbers on the abscissa represent respectively the quantities a and b defined in the text.*

The thermal resistance of a single thermocouple determines the total thermal resistance of the thermopile which contains thousands of thermocouples thermally in parallel. Therefore, the thermal resistance of each thermocouple should be increased by reducing the width and increasing the height. However, these structural parameters in the practical design are subject to the technology limit, such as the depth of focus in contact photolithography.

The TEG on human body has been modeled via a network of thermal resistors representing different components, such as radiator and human body. The output performance of a watch-size TEG on human body for various design types are shown in Fig. 9. Type A contains 2350 thermocouples while type B contains 4700 thermocouples. An output power of about 1 μW at an output voltage of more than 1 V is expected from the current design of thermopiles in a watch-size generator.

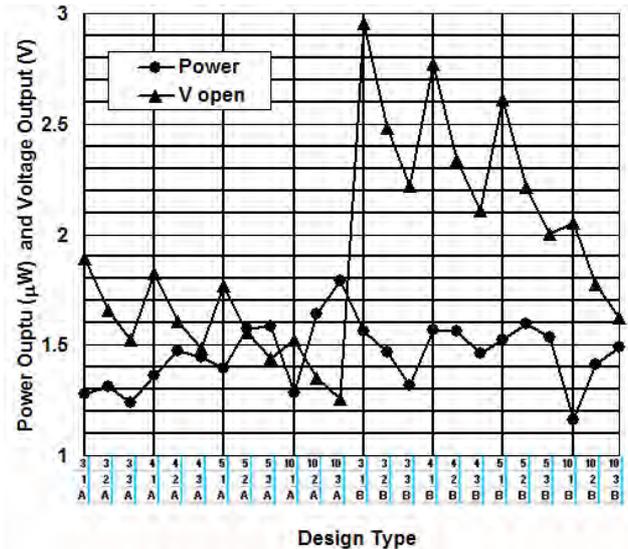

*Fig. 9. Calculated output voltage and power from various types of TEGs in the current design. . The three rows of numbers on the abscissa represent, from top to bottom, the quantities a and b defined in the text and the number of thermocouples (A= 2350 and B=4700).*

## 5. FABRICATION

The thermopile is fabricated using surface micromachining technology, as shown in Fig. 10. The Si substrate is firstly patterned and etched to form the trenches (Fig. 10(a)). Then the trenches, which are 2.5 μm deep, are filled with the first layer of sacrificial TEOS and the wafer is planarized by chemical mechanical polishing (CMP). In the next step, an additional layer of TEOS, which is 0.5 μm high, is deposited on top and patterned in order to form the step between cold and hot junctions. A thin layer of $Si_3N_4$ is then deposited for electrical insulation (Fig. 10(b)). P- and n-type poly-SiGe are deposited by LPCVD and patterned using dry etch to form thermocouples (Fig. 10(c)). An aluminum layer is sputtered and patterned to interconnect the two types of poly-SiGe legs (Fig. 10(d)). In order to form ohmic contact, an argon plasma clean is done before the sputtering and a sintering in forming gas atmosphere is performed afterwards. The rim is then formed by dry etch as deep as 250 μm (not shown) and the top die is bonded to the thermocouples (Fig. 10(e)). Finally, the device is released in a mixture of HF and glycerol and diced (Fig. 10(f)).

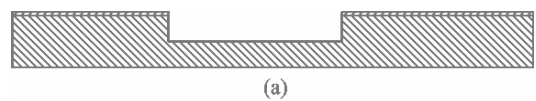

(a)

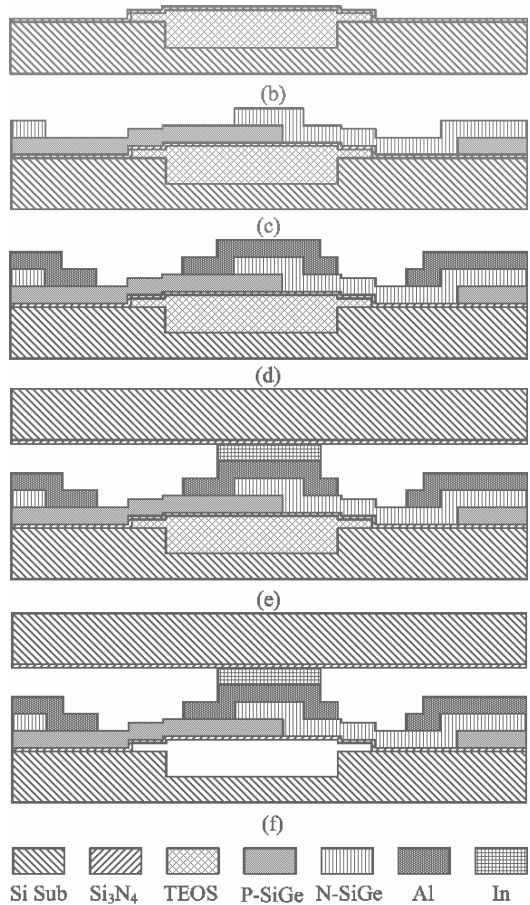

Si Sub | Si₃N₄ | TEOS | P-SiGe | N-SiGe | Al | In

*Fig. 10. Key process steps for the thermocouple array (a) pattern trench in a Si substrate; (b) deposit and pattern sacrificial TEOS layer; (c) deposit and pattern p- and n-type SiGe; (d) deposit and pattern Al; (e) bonding with top die; (f) final vapor HF release.*

The fabricated thermopile placed on the rim structure on the bottom chip is shown in Fig. 11.

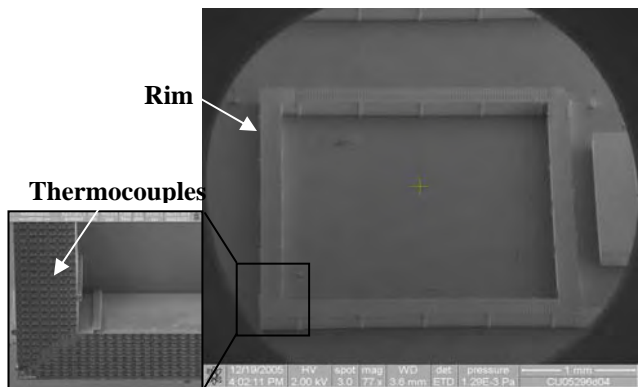

Rim

Thermocouples

*Fig. 11. SEM picture of the micromachined thermopile on the rim.*

The relevant material properties of p- and n-type poly-SiGe used in the thermopile have been characterized. The electrical resistivity is 1.05 mΩcm and 5.87 mΩcm for p-and n-type poly-SiGe respectively. The specific contact resistance between poly-SiGe and aluminum has been measured by cross bridge Kelvin structures. The specific contact resistance for p-type poly-SiGe is 86 $\Omega\mu m^2$ and that for n-type poly-SiGe is 40 $\Omega\mu m^2$. The Seebeck coefficient has been measured by a specially designed experimental set-up. The average Seebeck coefficient for p-type poly-SiGe is 69 µV/K and that for n-type poly-SiGe is -248 µV/K.

Although thermal conductivity has not been measured, a value of 3 W/K/m can be assumed, as measured on poly-SiGe deposited in the same conditions and in the same system before. Therefore, the thermoelectric figure-of-merit at room temperature, which characterizes the overall thermoelectric properties of materials, is 0.025 and 0.096 for p- and n-type poly-SiGe respectively. The improvement of the figure-of-merit of p-type poly-SiGe, which is relatively low, would lead to higher TEG output performance.

## 5. MEASUREMENT

The thermopile on the rim (Fig. 11) and the top die have been fabricated separately. Their assembling is ongoing at this moment. The functionality of the thermopile has been demonstrated successfully before the HF release. The thermopile has been heated up on a thermal chuck. The output voltage is measured as a function of the chuck temperature. In such conditions, the temperature drop on the thermopile would be a small fraction of 1 K. In order to increase the temperature gradient between the hot and cold junctions, a rudimental radiator, made of a bare silicon die, is positioned onto the thermopile touching the cold junctions only. The temperature difference can be further increased by flushing the measurement chamber with nitrogen.

The experimental results shown in Fig. 12 were obtained on a thermopile of the type 10-3-A. The output voltage increases with the temperature difference between the thermal chuck and the ambient air. Before the rim is made, the output voltage is about 13 mV/K/cm² without nitrogen flush. With the nitrogen flush, the output voltage increases by a factor of 4 and is as large as 53 mV/K/cm². After the rim is formed, the output voltage is then increased to about 30 mV/K/cm² and 130 mV/K/cm² respectively for the conditions without and with nitrogen flush. This improvement illustrates the advantage of the rim structure in increasing the temperature difference across thermocouple junctions.

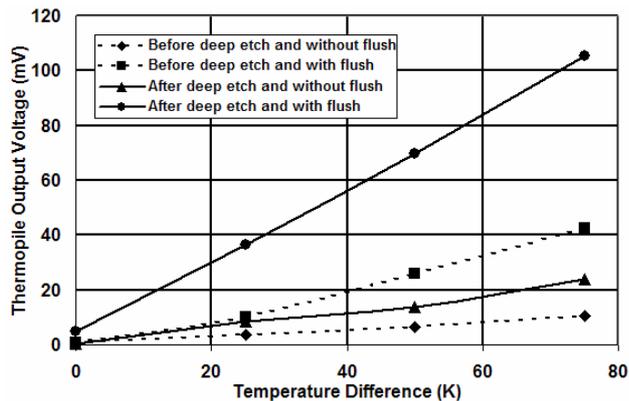

*Fig. 12. Dependence of the output voltage on the temperature difference between the chuck and the ambient air.*

A larger output voltage is expected after release of sacrificial TEOS because its removal improves the thermal isolation between the cold and hot junctions.

## 6. CONCLUSION

A micromachined poly-SiGe-based thermopile for human body applications has been designed and modeled. In order to improve the output performance, a rim structure has been introduced. The influence of structure parameters on the thermal resistance of single thermocouple has been investigated by FEM simulation. Analytical models show that about 1-2 μW at a voltage of more than 1 V can be obtained with the current thermopile design in a watch-size TEG when placed on human body. The thermopile has been fabricated and tested. Poly-SiGe used as the thermoelectric material has been characterized. The functionality of the thermopile has been demonstrated in thermal measurements before the final release. The increase in output voltage due to the rim structure has been observed. In presence of forced convection, the measured output voltage is increased from about 53 mV/K/cm$^2$ to about 130 mV/K/cm$^2$ after the rim structure is formed. A further increase in output voltage is expected upon completing the full device.

## 7. REFERENCES

[1] V. Leonov, T. Torfs, P. Fiorini and C. Van Hoof, "Thermoelectric Converters of Human Warmth for Self-powered Wireless Sensor Nodes", *IEEE Sensor Journal, Special Issue on Intelligent Sensors*, 2007 (in press).

[2] H. Böttner, J. Nurnus, A. Gavrikov, G. Kühner, M. Jägle, C., Künzel, D. Eberhard, G. Plescher, A. Schubert and K.-H. Schlereth, "New Thermoelectric Components Using Microsystem Technologies", in *IEEE Journal of MicroElectroMechanical Systems*, Vol.13, No.3, pp.414-420, Jun. 2004.

[3] J. P. Fleurial, G. J. Snyder, J. A. Herman, P. H. Giauque and W. M. Philips, "Thick-film Thermoelectric Microdevices", *18th International Conference on Thermoelectrics*, pp. 294-300, Baltimore, U.S., Aug. 1999.

[4] S. Hasebe, J. Ogawa, M. Shiozaki, T. Toriyama, S. Sugiyama, H. Ueno and K. Itoigawa, "Polymer Based Smart Flexible Thermopile for Power Generation", *17th International Conference on MicroElectroMechanical Systems*, pp. 689-692, Maastricht, the Netherlands, Jan. 2004.

[5] W. Glatz and C. Hierold, "Flexible Micro Thermoelectric Generator", *20th International Conference on MicroElectroMechanical Systems*, pp. 89-92, Kobe, Japan, Jan. 2007.

[6] M. Strasser, R. Aigner, C. Lauterbach, T. F. Sturm, M. Franosch and G. Wachutka, "Micromachined CMOS Thermoelectric Generator As On-chip Power Supply", *Sensors and Actuators A*, Vol.114, pp.362-370, 2004.